\documentclass{emulateapj}
\usepackage{amsfonts,amsmath,amssymb}
\usepackage{graphicx}
\usepackage{color}
\usepackage{url}

\usepackage[space]{grffile}
\usepackage{latexsym}
\usepackage{textcomp}
\usepackage{longtable}
\usepackage{multirow,booktabs}
\usepackage{natbib}
\usepackage{hyperref}
\hypersetup{colorlinks=false,pdfborder={0 0 0}}
\newif\iflatexml\latexmlfalse
\usepackage[utf8]{inputenc}
\usepackage[english]{babel}

\newcommand{\msun}{M_\odot}
\newcommand{\rxte}{{\it RXTE}}

\newcommand{\rin}{R_{\rm in}}

\newcommand{\rchinu}{\chi^{2}/\nu}
\newcommand{\fsc}{f_{\rm SC}}
\newcommand{\nh}{N_{\rm H}}

\newcommand{\keV}{\rm keV}

\newcommand{\cm}{\rm cm}

\newcommand{\HR}{\rm HR}

\newcommand{\ew}{{\rm EW}_{\rm PL}}
\newcommand{\EW}{{\rm EW}_{\rm PL}}

\newcommand{\xillver}{{\sc xillver}}

\begin{document}

\title{Stronger Reflection from Black Hole Accretion Disks \\
  in Soft X-ray States}


\shorttitle{Stronger Reflection in BH Soft States}
\shortauthors{Steiner et al.}

\author{James F.\ Steiner\altaffilmark{1}\altaffilmark{\dag}, 
Ronald A.\ Remillard\altaffilmark{1}, 
Javier A.\ Garc\'ia\altaffilmark{2}, 
Jeffrey E.\ McClintock\altaffilmark{2}}
\altaffiltext{\dag}{Einstein Fellow.}

\altaffiltext{1}{MIT Kavli Institute for Astrophysics and Space  Research, MIT, 70 Vassar Street, Cambridge, MA 02139.}
\altaffiltext{2}{Harvard-Smithsonian Center for Astrophysics, 60  Garden Street, Cambridge, MA 02138.}

\email{jsteiner@mit.edu}
\begin{abstract}
We analyze 15,000 spectra of 29 stellar-mass black hole candidates collected over
the 16-year mission lifetime of \rxte\ using a simple phenomenological
model. As these black holes vary widely in luminosity and progress
through a sequence of spectral states, which we broadly refer to as hard
and soft, we focus on two spectral components: The Compton power law and
the reflection spectrum it generates by illuminating the accretion
disk. Our proxy for the strength of reflection is the equivalent width
of the Fe-K line as measured with respect to the power law.  
A key distinction of our work is that for {\em all} states we estimate the
continuum under the line by excluding the thermal disk component and using
only the component that is responsible for fluorescing the Fe-K line,
namely the Compton power law.
We find that reflection is several times more pronounced ($\sim 3$) in soft compared to hard spectral states. This is most readily caused by the dilution of the Fe line amplitude from Compton scattering in the corona, which has a higher optical depth in hard states.   Alternatively, this could be explained by a more compact corona in soft (compared to hard) states, which would result in a higher reflection fraction.

 
\end{abstract}

\section{Introduction}\label{section:intro}

During the course of its 16-year mission, the {\it Rossi X-ray Timing
  Explorer} (\rxte) detected far more photons (30 billion in PCU-2 alone) from
accreting black holes than any other X-ray observatory. The sample of black holes (BHs) targeted by {\it RXTE} is chiefly comprised of nearby stellar-mass systems.   While the total Galactic population of stellar BHs is
believed to be many millions, only a tiny subset of approximately 50 are
known to us, namely those located in X-ray binaries.

A wondrous property of BHs, their utter simplicity, is the essence of
the famous no-hair theorem: Each BH in nature is fully described by
just its mass and spin. Roughly half of the known stellar BHs have a
dynamically-determined mass. The measured masses range from
$\sim5-20~\msun$ \citep{Ozel_2010,Reid_2014,Laycock_2015,Wu_2016}.
Meanwhile, estimates of spin have been obtained for many of them during
the past decade, principally by modeling either the thermal continuum
emission of the accretion disk \citep[e.g.;][]{Zhang,MNS14}, or the
relativistically-broadened reflection spectrum
\citep[e.g.;][]{Fabian_1989,Reynolds_2014}.

Our focus is primarily on transient BH systems that cycle between a
minuscule fraction of the Eddington limit upward to the limit
itself. During an outburst, a transient BH progresses through a sequence
of spectral-timing states, which are broadly termed ``hard'' or
``soft,'' based on a measure of X-ray hardness \citep{Fender_2004}. As a
source evolves over the course of months and its hardness varies,
sweeping changes occur in many properties of the system including the
composition of its spectrum, the intensity of Fourier flicker noise, and
the presence or absence of quasi-periodic oscillations and jets
\citep[e.g.;][]{Homan_2005,RM06,Heil_2015}.

Stellar BHs emit a complex multicomponent X-ray spectrum. A {\it
  thermal} blackbody-like component is produced in the very inner
accretion disk. The disk is truncated at a radius $\rin$ before
reaching the event horizon. A hard {\it power-law} component results
from Compton scattering of the thermal disk photons in hot coronal gas
that veils the disk. The third principal component is a {\it reflection}
spectrum generated by illumination of the cold disk ($kT\sim0.1-1$\,keV)
by the power-law component. The reflection component is a rich mix
of radiative recombination continua, absorption edges and fluorescent
lines \citep{Ross_1993,Garcia_Kallman_2010}. An analysis of these three
interacting spectral components provides constraints on the source
properties including geometry (e.g., on $\rin$ and the scale of the
corona). The relationships between these components across the full range
of behavior displayed by accreting stellar BHs is the focus of this
paper.

Our results are based on an analysis for 29 stellar BHs (10
dynamically-confirmed BHs and 19 BH candidates) of all the data
collected using \rxte's prime detector unit (PCU-2), some 15,000 spectra
in all, with a total net exposure time of 30\,Ms. Importantly, we
recalibrate the data using our tool {\sc pcacorr}, which greatly reduces
the level of systematic error \citep{pcacorr}. Given the scope of
our study, relativistic reflection models are too complex and
computationally slow for our purposes \citep[e.g.; {\tt reflionx,
  xillver, relxill;}][]{reflionx,relxill2}. We therefore employ a
simplistic, phenomenological model and estimate the strength of the
reflection spectrum by determining the equivalent width with respect to the Compton continuum of its most
prominent reflection feature, namely the $6.4-7.0$~keV Fe-K line.


The paper is organized as follows:  In Section~\ref{section:data} we describe
the data sample and our approach to modeling the data.  Our results are
presented in Section~\ref{section:results}, followed by a discussion in
Section~\ref{section:disc} and our conclusions in Section~\ref{section:conc}.

\begin{figure}[]
\begin{center}
\includegraphics[width=1\columnwidth]{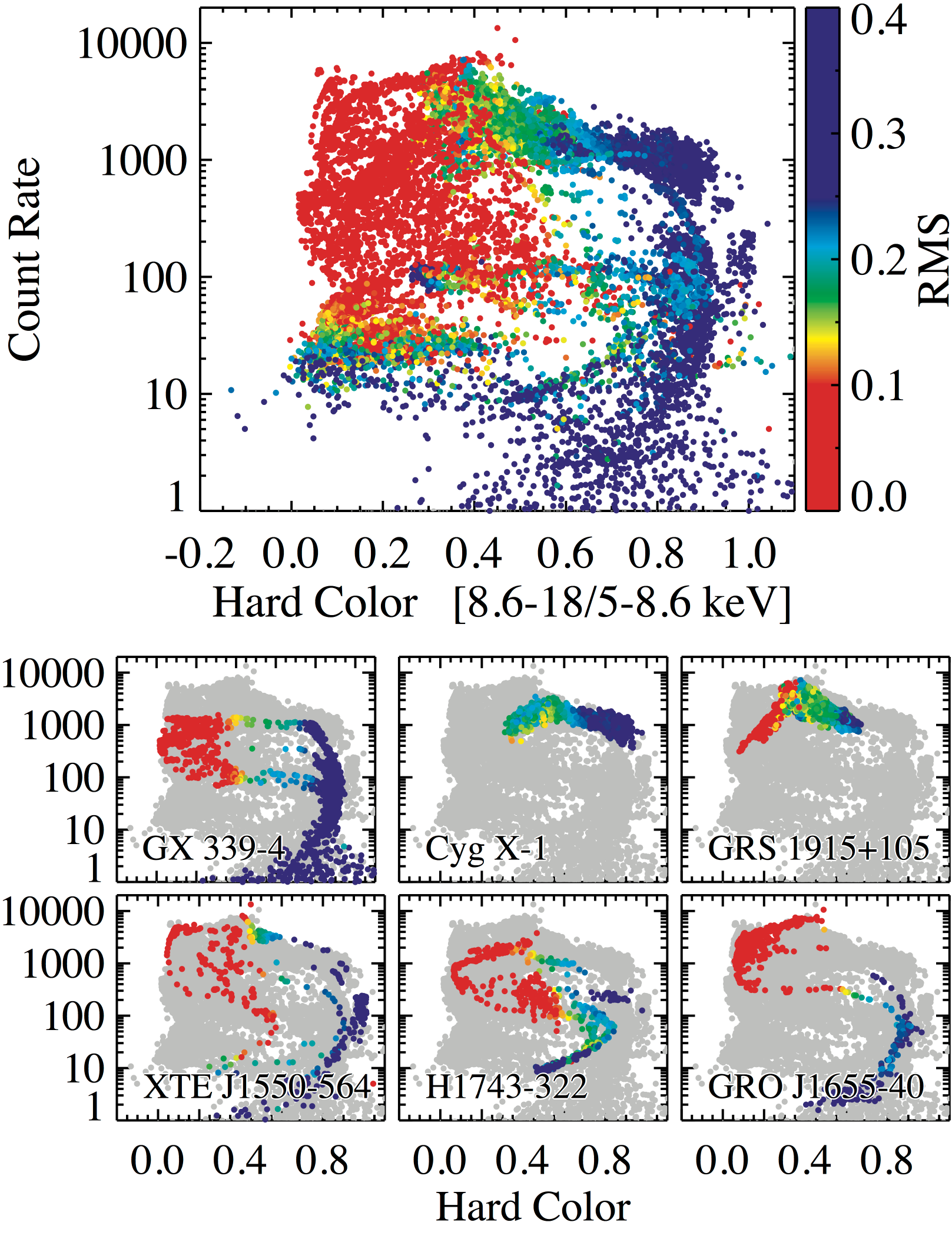}
\caption{({\it top}): Hardness-intensity diagrams for all data and ({\it
    bottom:}) for six well-known BHs with abundant data (where for
  reference the gray background shows all data).  For reference, the count rate of the Crab Nebula is $\approx
  2600$~s$^{-1}$. Note that a HID does not allow one to compare the
  luminosities of sources because the intensity is in detector
  units.}\label{fig:qdiag}
\end{center}
\end{figure}

\section{Data}\label{section:data}

The \rxte\ archive provides the premier database for the synoptic study
of stellar BHs. We exclusively use the data collected by PCU-2, one of
the five proportional counter detectors that comprise \rxte's principal
instrument, the Proportional Counter Array (PCA). Throughout the
mission, PCU-2 was the unit that was most often active, and it had the
most reliable and stable calibration \citep{Jahoda_2006,
  Shaposhnikov_2012}. Its area and energy resolution were 1300\,cm$^2$
and $\approx18$\% at 6 keV. Table~1 summarizes our data sample.

During an outburst, a BH was typically observed daily over a period of
months as it systematically brightened and subsequently dimmed by orders
of magnitude. We homogenized the data by segmenting it into continuous
300--5000\,s intervals, each of which was used to produce an energy
spectrum and a power-density spectrum (PDS). Energy spectra were
analyzed ignoring the lowest 4 channels, an effective lower bound $\approx
2.8$ keV, and an upper bound of 45\,keV was adopted. The effects of
detector dead time were corrected as described in
\citet{McClintock_2006}. We obtained an absolute calibration of the flux
using the the standard \citet{Toor_Seward} spectrum of the Crab Nebula;
our slope and normalization corrections are $\Delta\Gamma = 0.01$ and
$f_{\rm TS} = 1.097$ \citep{Steiner_2010}. We computed the rms power, a
measure of the flicker noise, by integrating the PDS over the band
0.1--10\,Hz.

An unprecedented sensitivity to faint spectral features is achieved by
employing the calibration tool {\sc pcacorr} \citep{pcacorr}, which
improves the quality of the PCA's spectral calibration by roughly an
order of magnitude and results in a data precision of $\sim 0.1\%$. We
include this small systematic uncertainty as a fractional error on each
channel when conducting our analysis using {\sc XSPEC}
\citep{XSPEC}. The considerable increase in sensitivity {\sc pcacorr}
delivers is crucial for estimating the strength of line features.

All PCU-2 data for 29 black holes are plotted in a 
hardness-intensity diagram (HID) \citep{Fender_2004,RM06} in the top
panel of Figure~\ref{fig:qdiag}.  The normalized hard color (or hardness ratio $\HR$) 
  is the ratio of count rates in the energy bands indicated in the upper
  panel, and is described in \citet{Peris_2016}.  The data are color-coded to show the
level of rms flicker noise. As is well-known and is evident here from the vertical striation,
rms noise correlates with spectral state
\citep[e.g.;][]{Heil_2015,RM06}, with hard states showing several-times
stronger rms than soft states.
The six small panels are HIDs for selected sources.
Note that transient sources characteristically trace a loop in the hardness-intensity diagram (HID), but that the persistent source Cyg X--1 is confined to a relatively narrow region.  The other selected source showing stunted HID evolution is GRS~1915+105, which is an unusual transient system that has been in a protracted state of outburst since 1992.

\subsection{Spectral Modeling}\label{subsec:model}

We adopt a single simplistic spectral model that is applicable to both
soft and hard spectra: {\tt
  phabs*[smedge(simpl$\otimes$diskbb)+gauss]}. The disk and Compton
components are modeled by {\tt diskbb} \citep{DISKBB}, and {\tt simpl}
\citep{Steiner_simpl}, respectively. The reflection component is
described by a Gaussian line with fixed energy of 6.5\,keV and an intrinsic width
of 50\,eV.  We note that owing to the broad detector resolution at the Fe line $\sim 1.2$~keV, the value adopted for the line width in our simplistic model is of minor consequence (as demonstrated in Section~\ref{section:disc}). 
  Despite its coarse resolution, \rxte\ is very sensitive to the line flux, as measured by the
  normalization of the Gaussian feature.  Accordingly, we adopt the line flux as our proxy
  for the intensity of the reflection component. We approximate the
  relativistically-broadened Fe-K absorption edge using {\tt smedge}
  \citep{SMEDGE}; we fit for the peak depth $\tau_{\rm smedge}$ with the
  width fixed at 7\,keV and the shape index set to -2.67
  \citep{Sobczak_2000}. Our adopted values of the column density $\nh$
  are summarized in Table\,1.


\begin{figure}[]
\begin{center}
\includegraphics[width=1\columnwidth]{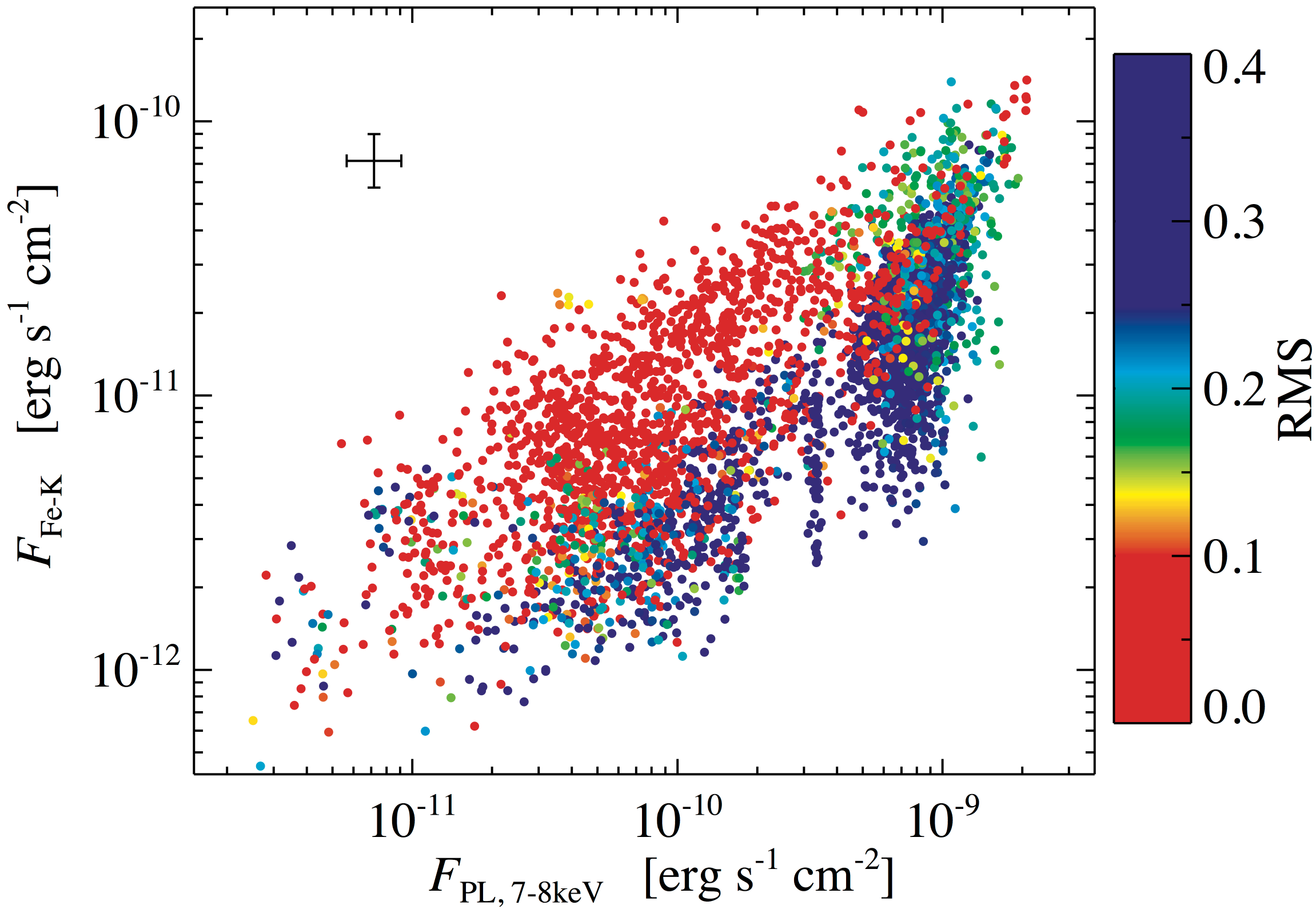}		
\includegraphics[trim={0cm 0cm 0
  -0.75cm},clip,width=1\columnwidth]{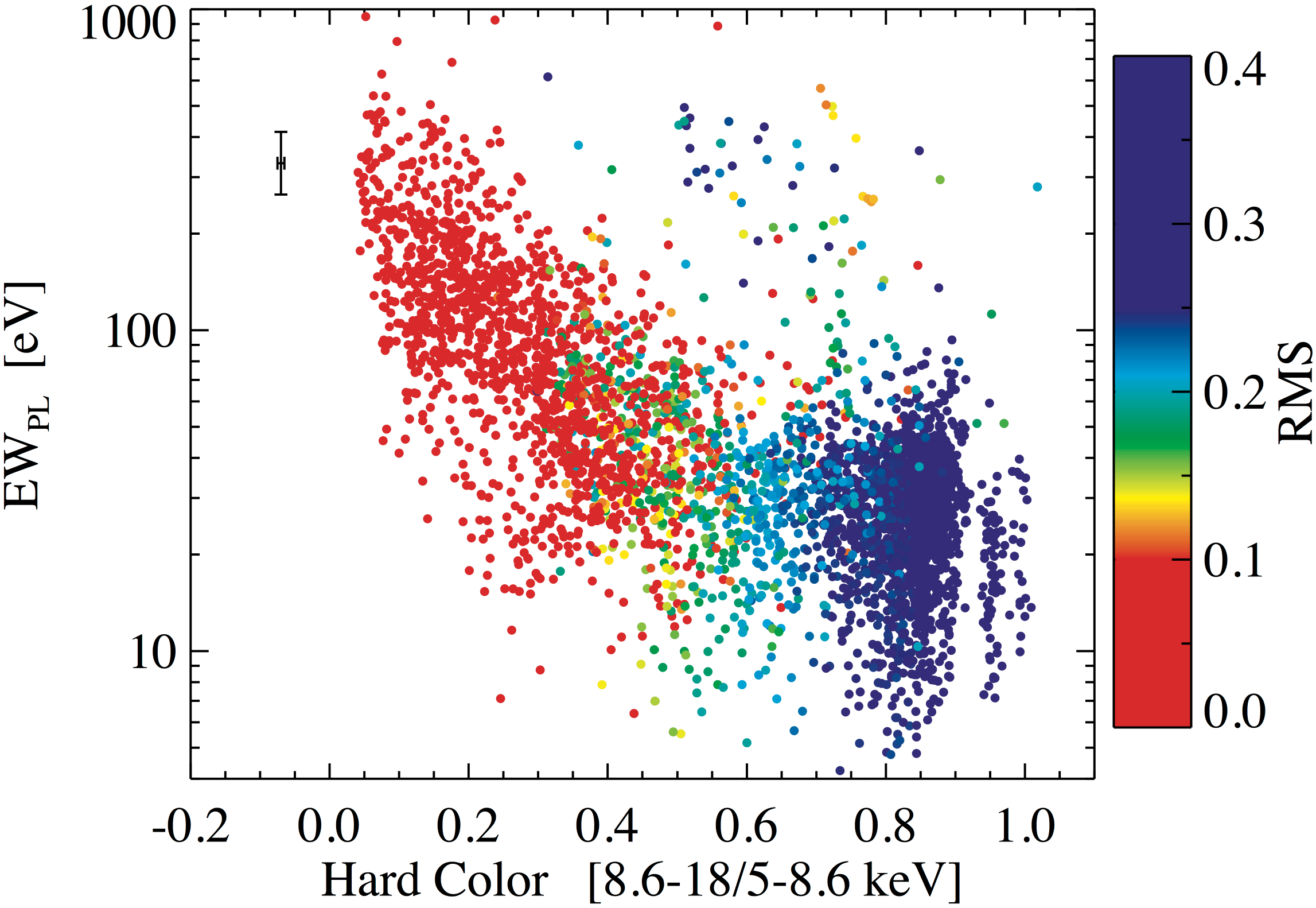}
\caption{({\it top}): Flux in the Fe line versus the adjacent flux in
the power-law continuum. Color-mapping indicates the rms power. ({\it
bottom}): Equivalent width of the Fe line computed using the
power-law flux only (i.e., excluding the disk flux). The reflection is
strongest in soft states, and it increases dramatically as the spectrum
softens.
Representative 1$\sigma$ error bars are shown in the upper-left
corner in each panel. 
}\label{fig:ew}
\end{center}
\end{figure}

\section{Results}\label{section:results}

We have fitted all $\approx 15,000$ spectra with our model.  
In Figure~\ref{fig:ew},
results are shown for approximately
one-quarter of these spectra, those that meet three criteria: $\rchinu <
2$; Fe-line normalization significant at the $\ge1\sigma$ level; and
power-law normalization $\fsc$ significant at the $\ge3\sigma$ level.
Plotted in the top panel is Gaussian line flux versus the Compton power
law flux near the Fe edge computed by integrating the power-law
component from 7--8\,keV.  Since the fluorescent Fe\,K line is produced
preferentially by photons at energies above and near the Fe\,K edge \citep{Reynolds_2009}, i.e., at energies $E\gtrsim7$\,keV 
\citep{Kallman_2004}, it is not surprising
that the line and continuum fluxes are strongly correlated. Though notably, this strong correlation is a direct validation of the reflection paradigm (wherein power-law emission fluoresces strong line emission).
What is surprising is that the soft states (red) appear to be much more efficient at
producing reflection than hard states (blue), as evidenced by the vertical offset between the clouds of soft and hard data.

This result runs counter to the conventional view that reflection is
weak in soft states (e.g., \citealt{Ross_2007, Yuan_2014}, but see the correlation between spectral index and reflection strength in \citealt{Zdziarski_1999}, and related work), a
view based on the weakness of reflection {\it relative to the thermal
  disk component}. Here instead, we appropriately relate the strength of
the reflection signal to the power-law component that produces it, as isolated from the evolving thermal disk.

In the lower panel of Figure~\ref{fig:ew}, we plot (versus the hardness ratio $\HR$) our proxy for the strength of reflection: i.e., we plot the equivalent width ($\ew$) of the
line with respect to the coronal flux. That reflection
is more pronounced for soft states is readily apparent. In soft states, $\ew$
decreases regularly by an order of magnitude as the spectrum hardens;
then, for intermediate and hard states it plateaus at $\HR \gtrsim
0.7$. Given the inhomogeneity of our data sample, which includes both
transient and persistent sources, the ordered quality of these data is
striking. 

\begin{deluxetable*}{lllrrrrr}
\label{tab:data}
\footnotesize
\tablewidth{0pt}
\tablecaption{{\it RXTE}'s PCU-2 BH Archive}
\tablehead{\colhead{System} & \colhead{RA} & \colhead{Dec} & \colhead{$N_{\rm obs}$} & \colhead{${\rm T_{obs}}$ (ks)} & \colhead{PCU-2 Counts ($10^6$)} & \colhead{$N_{\rm H} (10^{22}~\cm^{-2})$} & \colhead{Reference}}
\startdata 
LMC X-3 &           05 38 56.3 &     -64 05 03 &          704   &      1203.95   &               33.5     &  0.04  & [1] \\ 
LMC X-1 &           05 39 38.8 &     -69 44 36 &         1598   &      3120.72   &              106.4     &  0.7   & [2] \\ 
XTE J1118+480 &     11 18 10.8 &     +48 02 13 &          124   &       220.93   &               15.7     &  0.01  & [3] \\ 
GS 1354-64 &        13 58 09.9 &     -64 44 05 &           23   &        54.46   &                6.2     &  2     & [4] \\ 
4U 1543-47 &        15 47 08.6 &     -47 40 10 &          130   &       243.26   &              278.2     &  0.4   & [3] \\ 
XTE J1550-564 &     15 50 58.8 &     -56 28 35 &          517   &       970.78   &             1875.6     &  0.8   & [5] \\ 
4U 1630-47\tablenotemark{a} &  16 34 01.6 & -47 23 35 &  1194   &      2108.10   &             1034.0     &  11    & [6] \\ 
XTE J1650-500 &     16 50 01.0 &     -49 57 44 &          195   &       337.28   &              119.5     &  0.5   & [3] \\ 
XTE J1652-453 &     16 52 20.3 &     -45 20 40 &           61   &        99.84   &                9.2     &  6.7   & [7] \\ 
GRO J1655-40 &      16 54 00.1 &     -39 50 45 &          987   &      2483.87   &             5174.2     &  0.7   & [3] \\ 
MAXI J1659-152 &    16 59 01.7 &     -15 15 29 &           71   &       146.64   &               66.7     &  0.25  & [8] \\ 
GX 339-4 &          17 02 49.4 &     -48 47 23 &         1672   &      2868.46   &              894.2     &  0.3   & [9] \\ 
IGR J17091-3624 &   17 09 08   &     -36 24.4  &          256   &       479.60   &               46.8     &  1.2   & [10] \\ 
XTE J1720-318 &     17 19 59.0 &     -31 45 01 &          122   &       283.36   &               80.7     &  1.3   & [4] \\ 
GRS 1739-278 &      17 42 40.0 &     -27 44 53 &           12   &        26.45   &               15.2     &  3.7   & [4] \\ 
H1743-322 &         17 46 15.6 &     -32 14 01 &          649   &      1403.18   &              958.5     &  2.2   & [11] \\ 
XTE J1748-288 &     17 48 05.1 &     -28 28 26 &           39   &        98.58   &               46.6     &  7.5   & [4]  \\ 
SLX 1746-331 &      17 49 48.3 &     -33 12 26 &           78   &       186.19   &               36.2     &  0.4   & [3]  \\ 
XTE J1752-223 &     17 52 15.1 &     -22 20 33 &          234   &       435.52   &              147.0     &  0.6   & [12]  \\ 
Swift J1753.5-0127 &17 53 28.3 &     -01 27 06 &          376   &       853.95   &              123.4     &  0.15  & [3]  \\ 
XTE J1817-330 &     18 17 43.5 &     -33 01 08 &          191   &       430.11   &              270.7     &  0.15  & [4]   \\ 
XTE J1818-245 &     18 18 24.4 &     -24 32 18 &           56   &       141.74   &               19.7     &  0.5   & [13]  \\ 
V4641 Sgr &         18 19 21.6 &     -25 24 26 &           94   &       179.57   &                3.0     &  0.25  & [3]  \\ 
MAXI J1836-194 &    18 35 43.4 &     -19 19 12 &           76   &       124.83   &                9.4     &  0.15  & [3]  \\ 
XTE J1859+226 &     18 58 41.6 &     +22 39 29 &          170   &       336.24   &              270.3     &  0.2   & [14]  \\ 
GRS 1915+105 &      19 15 11.6 &     +10 56 45 &         2566   &      5255.58   &            12520.1     &  6     & [15]  \\ 
Cyg X-1 &           19 58 21.7 &     +35 12 06 &         2446   &      5413.02   &             6361.3     &  0.7   & [3]  \\ 
4U 1957+115 &       19 59 24.2 &     +11 42 32 &          243   &       646.42   &               41.0     &  0.15  & [3]   \\ 
XTE J2012+381 &     20 12 37.7 &     +38 11 01 &           30   &        53.98   &               13.5     &  1.3   & [4]  \\ 
\hline
Total &             \nodata &         \nodata &         14914   &       30207.   &             30577.     &  \nodata & \nodata \\  
\enddata

\tablenotetext{a}{For 76 observations the source was offset
 in the 1-deg FWHM collimator by $\sim0.6$\,degrees, which significantly
 reduced the count rate. No other source has more than several pointings
 offset by $> 0.5$ degrees.}
\tablecomments{ References: [1] \citealt{Steiner_2010};
 [2] \citealt{Gou_2009};
 [3] \citealt{DL90};
 [4] \citealt{Dunn_2010};
 [5] \citealt{Steiner_j1550spin_2011};
 [6] \citealt{Tomsick_2005};
 [7] \citealt{Hiemstra_2011};
 [8] \citealt{Yamaoka_2012};
 [9] \citealt{Hynes_2004};z
 [10] \citealt{Rodriguez_2011};
 [11] \citealt{JEM_H1743};
 [12]  \citealt{Nakahira_2012};
 [13]  \citealt{CadolleBel_2009}; 
 [14]  \citealt{Farinelli_2013};
 [15]  \citealt{Feroci_1999}}
\end{deluxetable*}

\subsection{A Case Study in Reflection: XTE J1550--564}\label{subsec:j1550}

To more precisely study the behavior of reflection in soft spectral
states, we examine one system in detail. We select an exceptionally
bright transient with abundant data: XTE J1550--564 (hereafter,
J1550; see the bottom-left subpanel of Fig.~\ref{fig:qdiag}). J1550 was
discovered on 1998 September 6, and two weeks later it reached a peak
intensity of 6.8\,Crab (2--10\,keV). Four additional fainter outbursts
were observed during the following decade.

\begin{figure}[]
\begin{center}
\includegraphics[width=1\columnwidth]{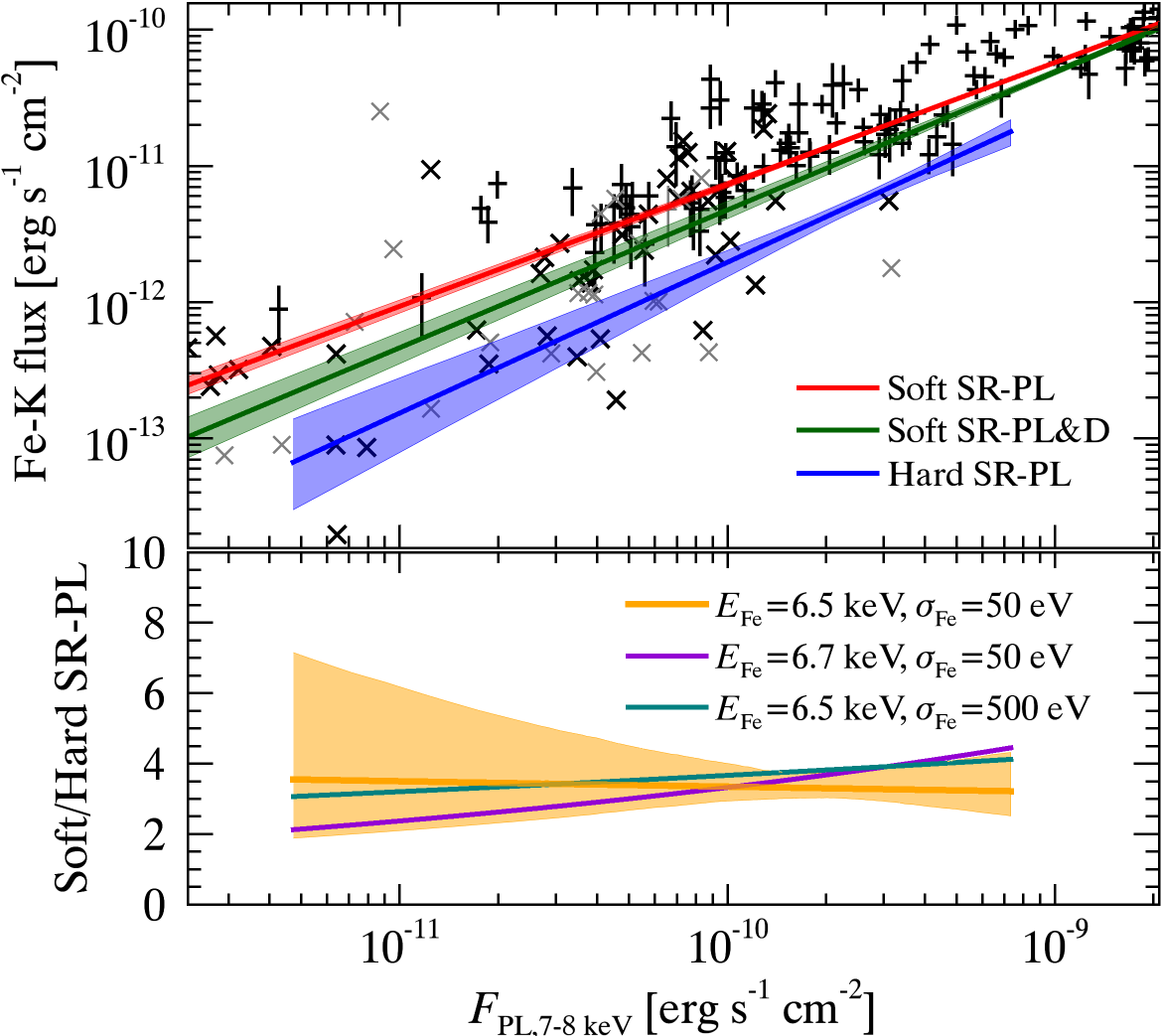}
\caption{({\it top}): Fe-line flux versus 7--8~keV power-law flux for J1550. Our
   primary ``SR-PL" scaling relation is shown in red with its associated 95\% confidence
   region. The ``SR-PL\&D" curve in green includes the effects of
   returning radiation. For reference, the SR-PL fit to the hard 
   data is shown in blue. At the level of detail, only those data with
   total flux $>3$\,mCrab
   and that yielded fits with $\rchinu < 2$ are
   shown.  For clarity, data are marked with crosses and error bars
   omitted when $|N_{\rm line}|/\sigma_{\rm line} < 2$.  Data consistent
   with absorption (i.e., $N_{\rm line} < 0$) are plotted in gray.
({\it bottom}):
Ratio of the Fe-line flux for soft states to that for hard
states computed using the SR-PL scaling relation. Our default curve is presented in
orange with its associated 95\% confidence intervals as a light-shaded region.
Results for alternate spectral models with different values of the Gaussian line
energy and line width, which produce modest systematic differences, are also
shown for comparison.}\label{fig:j1550}
\end{center}
\end{figure}

We define $\HR = 0.7$ as the cut between soft and
hard-state data, and we again (as in Figure~\ref{fig:ew}) examine the
relationship between our proxy for reflection (namely the flux in the Fe
K line) and the flux in the adjacent Compton continuum. The data are plotted in
Figure~\ref{fig:j1550}. 
We begin with a simplistic assumption that the coronal flux and Fe line flux will 
scale together, i.e., $F_{\rm Fe} \equiv \alpha F_{\rm PL,7-8~\keV}^\beta$, and we proceed to fit for $\alpha$ and $\beta$.  This {\em scaling relation} is termed SR-PL to emphasize that the line flux scales with the power law.  We likewise pursue the SR-PL fit for hard spectra (i.e., $\HR > 0.7$), with the constants of soft and hard data determined independently.

For the soft data, we additionally investigate the possibility that disk self-irradiation may also contribute to the reflection emission.  This is a motivated notion given that bright  soft-states  can produce appreciable thermal emission even above 5~keV, and further, some fraction of the thermal photons (the ``returning radiation'') is bent back and strikes the disk rather than escaping to infinity.  We consider returning radiation by allowing for an added scaling with the disk's emission in the same 7--8 keV band proximate to the Fe-K edge, i.e., $F_{\rm Fe} \equiv \alpha F_{\rm PL,7-8~\keV}^\beta + \gamma F_{\rm disk,7-8~\keV}^\delta$ (this is the SR-PL\&D scaling relation), and we likewise fit for its parameters.




In the top panel of Figure~\ref{fig:j1550}, we show the Compton component's contribution to the Fe-line flux for both
the SR-PL (red) and SR-PL\&D (green) fits. The contribution
to the line flux from returning radiation is vanishingly
small, particularly at the highest luminosities.  
In the same figure, for reference we also show the SR-PL fit to the hard data in blue. Note that the
error bounds for both soft-data fits lie well above the hard correlation bounds. 

 The best-fitting scale indexes and 1$\sigma$
   errors are $\beta_{\rm SR-PL,soft}=0.90\pm0.01$, $\beta_{\rm
     SR-PL,hard}=1.12\pm0.12$, and $\beta_{\rm SR-PL\&D,soft}=1.01\pm0.03$,
   $\delta_{\rm SR-PL\&D,soft}=0.43\pm0.03$. The SR-PL\&D fit slightly
   outperformed the SR-PL fit ($\Delta\chi^2 \approx 20$ for 2 added
   degrees of freedom). 

Having established that the effects of returning radiation are minor, we
focus on comparing reflection for soft and hard data using solely the
SR-PL curves. We produce
Markov-Chain Monte Carlo realizations for both the hard and
soft SR-PL correlations (top panel, Figure~\ref{fig:j1550}), and we compare  
their respective Fe-line fluxes. The results are
shown in the bottom panel of Figure~\ref{fig:j1550}.  On average, the
Fe line is $>3$ times stronger for soft data, while the 95\%
confidence region ranges from $\sim 2-9$ times stronger.  
 
\section{Discussion}\label{section:disc}

As illustrated in Figure~\ref{fig:j1550}, our estimate of the ratio
of the Fe-line flux in soft and hard states is only modestly sensitive
to the values we adopted for the  line energy (6.5\,keV) and width
(50\,eV). Varying these values does not affect our conclusions. We have
also explored other model formulations (e.g., including a power-law
cutoff and replacing the Gaussian by a relativistic line profile) and
have similarly found that our conclusions are unaffected. Varying the
line shape systematically rescales the line flux, but it has a minor affect
on the ratio of the fluxes in soft and hard states, which is our focus. 

Consideration of two shortcomings of our simplistic model serve to
strengthen our conclusion that soft states are more efficient in
producing reflection emission. First, soft-state disks are hotter and
more strongly ionized, and hence they generally produce more reflection continuum
emission, which gets lumped together with the Compton power law. In
the case of our simple model, this effect serves to boost the power-law
continuum, thereby reducing $\ew$ for the soft state.
Secondly, Fe-line absorption features in disk winds 
are preferentially and often observed in soft
states \citep{Ponti_2012}, and these absorption features act to weaken the
emission line.  Our finding of enhanced $\ew$ in soft states is thus contrary to these biases.

\begin{figure}[]
\begin{center}
\includegraphics[trim={0cm 0cm 0
  -0.75cm},clip,width=1\columnwidth]{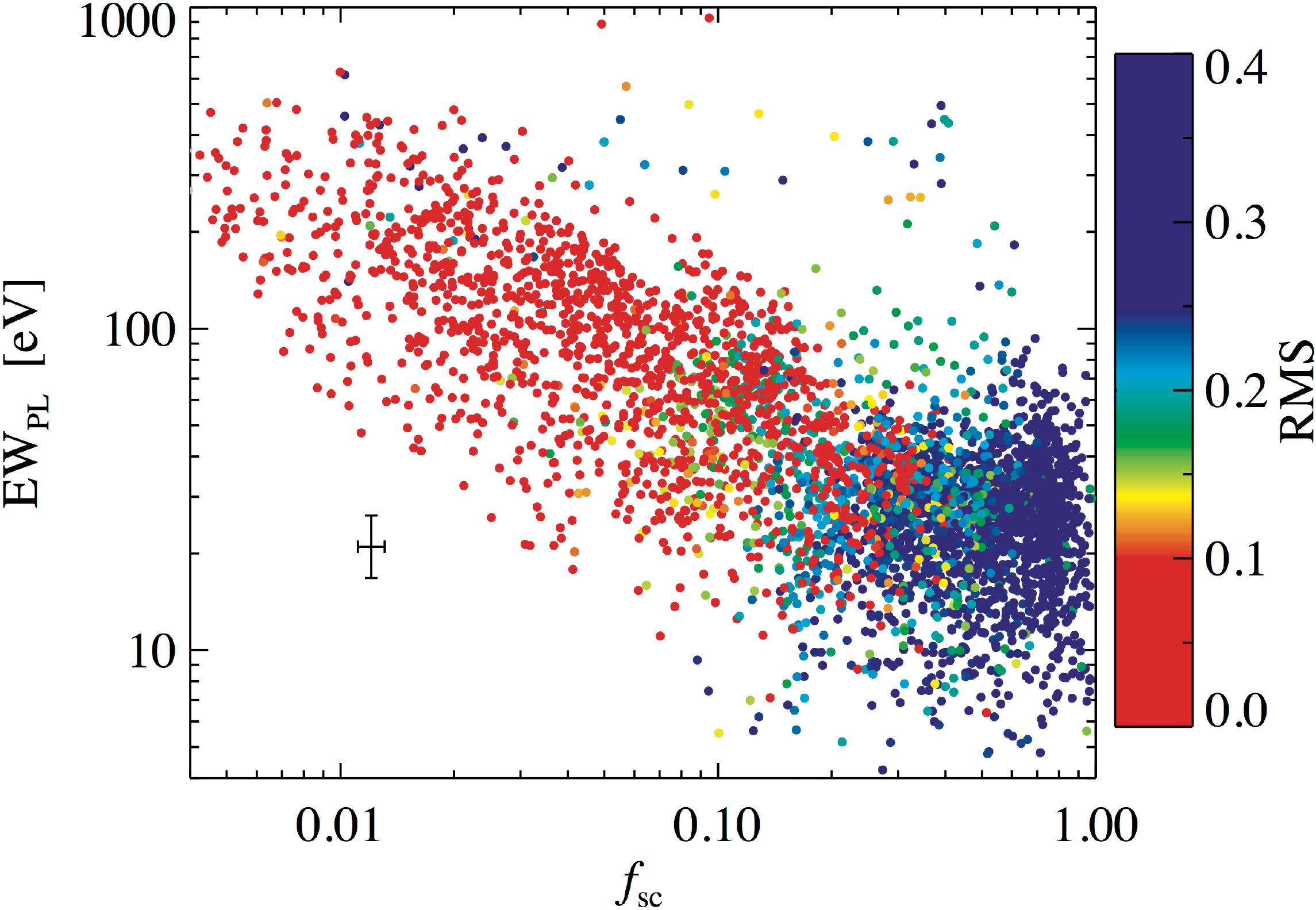}
\caption{Strong anti-correlation between $\fsc$ and $\EW$, as predicted
by \citet{Petrucci_2001}.  The dilution of line flux by Compton scattering 
of varying optical depth naturally accounts for the more pronounced $\EW$ in soft states.
}\label{fig:fsc}
\end{center}
\end{figure}

Earlier work by \citet{Petrucci_2001} which examined the effect of Comptonizing reflection spectra showed that at a fixed value of $\Gamma$, the fitted Fe-K equivalent width is very sensitive to the coronal temperature $kT_{\rm e}$, principally because as  $kT_{\rm e}$ decreases, optical depth $\tau$ necessarily increases (a straightforward consequence of the equations governing thermal Comptonization).  Accordingly, at high optical depth, the line is highly scattered in
the corona.  The scattered portion blends with the continuum, which results in a decrease in the Gaussian's equivalent width.  We have examined the evolution of equivalent width versus spectral hardness (changing $\Gamma$) and other spectral parameters. There is a strong anti-correlation between $\fsc$ and $\EW$ in precisely the sense predicted by \citet{Petrucci_2001}, and is the most apparent explanation for the observed trend.  This is shown in Figure~\ref{fig:fsc}.



Alternatively, or in addition to the dilution of the line by Compton scattering in the corona, changes in the disk-coronal geometry can impact the reflection's strength. In particular, a corona at very small scale-height ($h_{\rm corona}$) coupled with a close-in disk yields a higher reflection fraction than when either (1) the disk is truncated or (2) the corona is very large compared to the event horizon \citep{Dauser_2014}.  Gradual evolution of accretion flow with more compact disk-coronal geometry (i.e., lower $h_{\rm corona}/\rin$) in soft states would similarly contribute to the observed correlation.  

We caution that the role of ionization is complex and varies nonlinearly as other coronal attributes change, including most notably, $\Gamma$.  These differences are  principally related to changes in the atmosphere's temperature structure.  
For the range of ionization parameters observed around active stellar-mass BHs, i.e., log~$\xi \approx 2-4$, we simulated PCU-2 data using the \xillver\ reflection model \citep{Garcia_Kallman_2010}, and applied our analysis method.
We found that $\EW$ varies by a factor of $\sim 2-3$ over the span of $\Gamma$ ($1.4-3.4$) and $\xi$; this is insufficient to account for the full trend in Figure~\ref{fig:ew}.  Moreover, for log~$\xi \leq 3$, higher $\Gamma$ tends to produce lower $\EW$ whereas at log~$\xi \gtrsim 3$, the trend reverses.



%

As summarized by \citet{Zdziarski_2003}, and also \citet{Gilfanov_2014},
there is abundant evidence for a strong positive correlation between spectral
index and reflection strength 
(the ``$R-\Gamma$" correlation).  This has
been seen in both BH X-ray binaries and in AGN.  However, such work has been
largely confined to examination of hard states.
To our knowledge, our work is the first compelling evidence that the
correlation between spectral softness (increasing $\Gamma$) and reflection fraction continues and is
strongly amplified in soft, thermal states (i.e., $\Gamma \sim 2-3$).  Further,
from considering the \rxte\ archive of active BHs, as is readily apparent in
Figure~\ref{fig:ew}, this change is gradual and orderly amongst the full cast
of stellar BHs.
%


\section{Conclusions}\label{section:conc}

We have examined the strength of reflection in a global study of stellar BHs
using a simplistic, phenomenological spectral model.  We directly validate the
reflection paradigm, wherein power-law flux induces reflection emission.  In
separating possible contribution from disk self-irradiation, we demonstrate
that the power law's contribution is dominant.

Most importantly, we show that the corona produces reflection features up to an
order of magnitude more pronounced in soft rather than hard states.  The data
suggest an ordered transition in which the line-to-continuum strength declines gradually with spectral hardness.   This is the first time the ``$R-\Gamma$'' correlation has been shown to extend through (and increase in) BH soft states.
One possible explanation is that a more compact disk-coronal geometry in soft states would produce the observed trend.  However, the most natural explanation for this trend is suggested by \citet{Petrucci_2001}, who describe the dilution of line features emitted by the disk due to Compton-scattering in the corona.  In our case, because hard states have corona with higher optical depth than soft states, their line features are correspondingly weakened resulting in the observed anti-correlation between $\HR$ and reflection strength.  

\acknowledgments 

JFS has been supported by the NASA Einstein Fellowship grant PF5-160144. \\

{\it Facility:} \facility{RXTE}


%

\end{document}